\documentclass[a4paper, 12pt]{article}
\usepackage[dvips]{graphics}

\begin{document}

\title{\huge  CAN EXPERIMENTS DISTINGUISH BETWEEN RADIATIVE ELECTROWEAK SYMMETRY BREAKING
AND CONVENTIONAL BREAKING.}

\author{ F.  N.  Ndili\\
Physics Department \\
University of Houston, Houston, TX.77204, USA.}

\date{August, 2007}

\maketitle

\begin{abstract}
We present a comparative study of radiative electroweak
\\ symmetry breaking and  conventional standard model breaking, and
pose  the question whether experiments can  distinguish  one
breaking mode from the other. The importance of the problem lies
in the fact that the two breaking modes  have very different
physical interpretations concerning the mechanism of spontaneous
electroweak symmetry breaking and the origin of mass.

\end{abstract}

E-mail: fndili@uh.edu

\newpage

\section{INTRODUCTION}
Coleman and Weinberg  pointed out several years ago [1] that
quantum loops and radiative corrections can play an important role
in determining the structure of the vacuum of a quantum field or
system of fields and particles.  Specifically they showed that
while a  scalar field  with mass parameter $\mu^2 =0 $ and
classical tree potential is unable to develop a vacuum expectation
value (VEV) and  spontaneous symmetry breaking (SSB), such VEV and
SSB become achievable when we take quantum loops into account.
This is particularly so if the scalar field is coupled to a gauge
field and dimensional transmutation can occur. The overall
conclusion is that an alternative electroweak symmetry breaking
(EWSB) mechanism exists along side  the standard model EWSB
mechanism that is based on the classical tree potential:
\begin{equation}\label{eq: ndili1}
V(\Phi) = \frac{1}{2}\mu^2 \Phi^2  + \frac{\lambda}{4} \Phi^4
\end{equation}
with a non-zero negative mass parameter $\mu^2 < 0$.  The
Coleman-Weinberg alternative model dispenses with the mass
parameter $\mu^2 = 0$ in equation (1) but adds  quantum loops  or
higher order potential terms. This alternative mechanism has
become known as radiative Electroweak symmetry breaking (REWSB).    \\

The fact is that these two mechanisms do not just differ in what
terms appear or do not appear in the scalar field   potential, but
lead to very different physical perceptions of what causes
electroweak symmetry breaking.  The standard model based on
equation (1) with its $\mu^2 < 0$ attributes electroweak symmetry
breaking to a primordial fundamental scalar field that unlike any
existing particle known so far has a negative mass parameter
$\mu^2 < 0$. In contrast, the Coleman-Weinberg REWSB mechanism
would attribute EWSB to quantum dynamics and a quantum origin.
Such quantum dynamics if favored would widen the scope for our
search and understanding of EWSB and the origin of mass.  Another
way of looking at the matter is to say that in the standard model
mechanism CEWSB,  the quantum system starts off with a
pre-assigned mass scale  $\mu \ne 0, $ compared to the
Coleman-Weinberg  model REWSB where the same quantum system is
left to fix its own  scale of dynamics. The question arises what
role the pre-assigned scale parameter  $\mu$ plays in the quantum
dynamics, and whether we can find any observable features that
distinguish the outcome of these two dynamics.  \\

Because of the widely differing physical interpretations of the
two mechanisms, we consider it worthwhile to examine in some
quantitative detail, what observational features and signatures
distinguish REWSB on the one hand, from the conventional standard
model electroweak symmetry breaking (CEWSB) on the other hand.
Elias et. al. [2-6] have considered the problem from one perspective.\\

Our approach to the problem is to first set out in section 2,
aspects of electroweak symmetry breaking that do not depend on any
explicit choice of the scalar field potential $V(\Phi)$. Such
aspects will therefore be common to both REWSB and CEWSB and we
may call these aspects the core of electroweak symmetry breaking.
Thereafter, we take up equation (1) in section 3,  with its $\mu^2
< 0$, as one explicit choice of the scalar potential.  Then we
work out what new relations besides the core EWSB equations,
follow from this one choice of $V(\Phi)$. These new relations or
features we can call the specific signatures of the standard model
CEWSB. In section 4, we make a different  choice of the scalar
potential, by setting $\mu^2 = 0$ which is the Coleman-Weinberg
REWSB  model, with quantum loops added. We work out the new
features in relation to the core equations,  and call these the
signatures of REWSB. We proceed further in sections 5 and 6, to
re-examine the same REWSB using the more modern aspects of the
Coleman-Weinberg model known as the renormalization group improved
effective potential [7-10]. Further improvement using 2-loop
$\beta $ and $\gamma $ functions is considered in section 7. The
two sets of signatures, CEWSB and REWSB, are  compared in section
8. Our final results and conclusions are stated in section 9.  \\

\section{Core relations of electroweak symmetry \\ breaking}
In terms of a scalar field, the core features that break
electroweak symmetry even when no explicit scalar potential is
specified, are  a complex scalar field that may or may not be a
fundamental field.  This complex scalar field must be a doublet
under the electroweak $SU(2)_L \times U(1)_Y$ gauge symmetry. The
complex doublet scalar field must have  a non zero vacuum
expectation value we take to be v. Given such a field one
is able to immediately write or parameterize it  in the form:\\
\begin{equation}\label{eq: ndili2}
\Phi =  \left(
      \begin{array}{c}
      \Phi^+ \\ \Phi^o
      \end{array}
      \right)
  =  \left(
      \begin{array}{c}
      \phi_1^+  +i\phi_2^+ \\ \phi_3^o + i \phi_4^o
      \end{array}
      \right)
  = e^{i\xi(x).\tau/v}\left(
      \begin{array}{c}
      0 \\ \frac{v + h(x)}{\sqrt{2}}
      \end{array}
      \right)
\end{equation}
in which only one component field $\phi_3^o$ acquires the non-zero
vacuum expectation value written:
\begin{equation}\label{eq: ndili3}
 \langle \Phi \rangle = \langle \phi_3^o\rangle = \frac{v}{\sqrt 2} \\
\end{equation}
This form of the scalar field is all that is required to derive
many results and features of the EWSB. These several features
include various charged and neutral currents as well as couplings
of fermions and gauge bosons to the scalar field. The details of
these core features can be seen in several places such as [11].
Among these relations that we obtain without our specifying any
explicit scalar potential, are the gauge boson masses :
\begin{equation}\label{eq: ndili3}
M^2_{W} =  \frac{g_2^2 v^2}{4}
\end{equation}

\begin{equation}\label{eq: ndili4}
M^2_{Z} = \frac{v^2(g_2^2 +  g_1^2)}{4}
\end{equation}
where $g_2$ is $SU(2)_{L}$ coupling constant, and $g_1$ is
$U(1)_{Y}$  coupling constant. \\

It turns out also that a definitive  numerical value of v = 246
GeV can be obtained  without recourse to any explicit choice of
the scalar potential. One simply combines  the V-A current
structure of the gauge theory [11], with such accessible processes
as $\mu$ decays, and obtains expression for v in terms of the  Fermi constant :    \\
\begin{equation}\label{eq: ndili15}
 v  =  2^{-1/4}G_F^{-2} = 246 GeV.
\end{equation}
Given this value of v and the known masses of the gauge bosons
stated in equations (4) and (5), one deduces the values of the two
gauge coupling constants at the EWSB scale v:
\begin{equation}\label{eq: ndili15A}
 g_2 = 0.6585;   g_1 = 0.3407
\end{equation}

 Similarly from Yukawa couplings of equation (2) to fermions,
 we obtain  fermion mass relations and couplings:
\begin{equation}\label{eq: ndili5}
m_{f} = \frac{g_f v}{\sqrt 2}
\end{equation}
where  $g_f$ stands for  Yukawa coupling constant of
 a given fermion f, such as the top quark $g_t$.
 Again these  relations do not require any explicit choice
 or specification of the scalar potential. In particular,
 the relations hold whether $\mu = 0$ as in REWSB, or
 $\mu \ne 0$ (CEWSB). \\

We now note one area where we are not able to obtain any
information at all based on equation (2) alone, without specifying
some explicit form of the scalar potential.  This area is in the
determination of physical Higgs particle mass $m_h$, and the self
coupling parameter $\lambda$  of the scalar field and its
components. This becomes the area where we can focus our search
for any distinguishing features between REWSB and CEWSB. We will
therefore proceed by devising ways to determine these two
quantities $m_h$ and $\lambda$, first in a CEWSB model and then in
a REWSB model. Both quantities turn out to be derivable from some
explicit form of the scalar potential we now write as $V(\phi) $
where $\phi$ stands for the scalar field (component) that has the
non-zero VEV. Our main focus is in the structure and analysis of
such two potentials $V_{CEWSB} $ and $V_{REWSB}$, and
the Higgs masses and couplings  they predict. \\

\section{Conventional Standard  Model  Potential CEWSB}

The conventional EWSB is based on equation (1). This potential
defines the ground state of the system
\begin{equation}\label{eq: ndili8}
\frac{dV(\phi)}{d\phi(x)} = 0
\end{equation}
and gives a value of the  VEV of $\phi$ in terms of the parameters
of the chosen potential :
\begin{equation}\label{eq: ndili9}
\langle \phi(x) \rangle = \frac{v}{\sqrt 2} =
\sqrt{\frac{-\mu^2}{\lambda}}
\end{equation}
with $\mu^2  <  0.$ Upon plugging equations (2) and (10) into
equation (1), we obtain a mass relation for the physical Higgs
particle h(x):  It is :
\begin{equation}\label{eq: ndili10}
m_{h}^2 = - 2 \mu^2
\end{equation}
More formally  the mass of the Higgs particle is given in terms of
 a chosen potential by:
\begin{equation}\label{eq: ndili11}
\frac{d^2V(\phi)}{d\phi^2} |_{\phi = v/\sqrt 2}  =  m_{h}^2
\end{equation}

Equations (10) and (11) become new relations  which used
separately or in combination with the core equations (2) - (8) may
lead to specific signatures of the CEWSB. Thus from equations (10)
and (11) we get
\begin{equation}\label{eq: ndili12}
m_{h}^2 = - 2 \mu^2 = \lambda v^2
\end{equation}
as a relation specific to CEWSB model.  We can go further to use
the core value  v = 246 GeV and relate $m_h$ specifically to
$\lambda$ as:
\begin{equation}\label{eq: ndili13}
m_{h} = 246 \sqrt \lambda
\end{equation}
If we assume that the scalar field interaction was perturbative  \\
in the dynamics leading to EWSB  at scale v,  and that $\lambda
\le 1 $, we obtain that : $m_h \le 246GeV$ which becomes a
signature of CEWSB mechanism.  On the other hand, if $\lambda > 1
$, then CEWSB predicts $m_h > 246 $ GeV.  Notably however, the
CEWSB does not give us any specific value for Higgs mass nor a
value for $\lambda$.

\section{The Coleman Weinberg REWSB Potential}

 We consider next the Coleman Weinberg REWSB
potential and what signatures we can tag onto it. For a $\lambda
\phi^4 $ scalar field theory, Coleman and Weinberg obtained the
general expression for its effective potential:
\begin{equation}\label{eq: ndili120}
V_{eff}(\phi_c) =  - \sum_{n}^\infty \frac{1}{n!}\Gamma^{(n)}
(00....0)[\phi_c(x)]^n
\end{equation}
where $\phi_c$ is the classical  field that has non-zero vacuum
expectation value. $\Gamma^{(n)}(00...0)$ are 1PI vertex Greens
functions to which n external legs $[\phi_c(x)]^n $ attach, each
carrying zero momentum. When these 1PI functions are expanded
perturbatively into Feynman loops, the above series can be
re-organized into a perturbative loop expansion for the same
effective potential:
\begin{equation}\label{eq: ndili14}
V_{eff}(\phi) = V_o  +  V_{1L}(\phi)  +  V_{2L} (\phi)  +  .....
\end{equation}
where $V_o$ is the tree potential given by equation (1); $V_{1L}$
is the one-loop potential; $V_{2L}$ is the 2-loop potential,  etc.
The above loop potentials require in general to be renormalized.
The renormalized effective potential determines  the state of EWSB
and the Higgs mass through the standard equations:
\begin{equation}\label{eq: ndili1a}
\frac{dV(\phi)}{d\phi} |_{\phi = v/\sqrt 2}  =  0
\end{equation}
\begin{equation}\label{eq: ndili1}
\frac{d^2V(\phi)}{d\phi^2} |_{\phi = v/\sqrt 2}  =  m_{h}^2
\end{equation}
\begin{equation}\label{eq: ndili19}
\frac{d^4V(\phi)}{d\phi^4} |_{\phi = v/\sqrt 2}  = 6\lambda
\end{equation}

For a purely scalar field theory, the above loops involve only
virtual scalar particles. When however other fields are present in
the system that can couple radiatively to the  scalar field, the
loop contributions of these other fields to  $V_{eff}(\phi_c)$
have to be taken into account.  In the specific case of EWSB where
fermions and gauge bosons are present besides the scalar field,
the Coleman Weinberg potential generalizes to include  quarks,
leptons and gauge boson loops. These extended loop calculations
have been carried out in a number of  places [7, 11].  We write
down some of the results up  to 1-loop potential for REWSB. In one
form Cheng and Li write [11]:
\begin{equation}\label{eq: ndili21}
V_{eff}(\phi) = V_o  +  V_{1L}(\phi) = \frac{\lambda}{4} \phi^4 +
C (log \frac{\phi^2}{M^2} + ....)
\end{equation}
where $M^2$ is the renormalization scale; and C is given by:
\begin{equation}\label{eq: ndili22}
C =  \frac{1}{64 \pi^2}[3\sum_v m_v^4 + m_s^4 - 12\sum_f m_{f}^4]
\end{equation}
Here $m_{v}$ are vector boson masses ($W^+, W^-, Z^o)$;  $m_{s} =
m_h $ is scalar (physical Higgs) boson mass, and $m_{f}$ are the
 fermion masses, quarks and leptons. \\

 Using core equations (4)-(8) without using any equations from CEWSB section 3,
 we can rewrite  equation (21) in terms of coupling constants  as follows:
\begin{equation}\label{eq: ndili23}
C =  \frac{\phi^4}{64 \pi^2}\left[\frac{3(3g_2^4 + 2g_2^2g_1^2 +
g_1^4)}{16} +  12 \lambda^2 - \sum_f 3g_f^4 \right]
\end{equation}

If we  neglect contributions from all leptons and  quarks except
the top quark, we can write the  REWSB potential equation (20) up
to 1-loop potential  finally as [2] :
\begin{equation}\label{eq: ndili16}
V_{eff}(\phi) = \frac{\lambda \phi^4}{4} + \phi^4 \left[
               \frac{12 \lambda^2 - 3 g_t^4}{64\pi^2} +
               \frac{3(3g_2^4 + 2g_2^2 g_1^2 + g_1^4)}{1024 \pi^2}
               \right] (log\frac{\phi^2}{M^2} - \frac{25}{6})
\end{equation}
Here $g_t$ is the top quark loop Yukawa coupling.  The neglect of
other fermion couplings follows from equation (8),
 where a fermion Yukawa coupling is  proportional to its mass,
and the top quark with its dominant mass clearly overshadows all
other fermions. Regarding the  parameter M in the above equation,
we note that in the absence of a pre-set mass scale $\mu^2 = 0$ in
the REWSB system, the quantum dynamics sets its own scale which we
take to be the renormalization scale M, as well as the scale of
any SSB in the system.   Therefore we can re-define our
renormalization conditions on the effective potential as ;
\begin{equation}\label{eq: ndili17}
 \frac{dV(\phi)}{d\phi}|_{\phi = M} = 0
\end{equation}

\begin{equation}\label{eq: ndili1a}
\frac{d^2V(\Phi)}{d\Phi^2} |_{\phi = M}  =  m_{h}^2
\end{equation}
\begin{equation}\label{eq: ndili19}
\frac{d^4V(\Phi)}{d\Phi^4} |_{\phi = M}  = 6 \lambda
\end{equation}
In place of equation (23), we can also write the 1-loop effective
potential in the form due to  Ford et.al. [8],  and to Casas et.
al.[9-10].
\begin{eqnarray}\label{eq: ndili33}
V_{eff}^{RG} &=& \frac{6g_2^4\phi^4}{1024\pi^2} \left(
log\frac{g_2^2\phi^2}{4M^2} - \frac{5}{6} \right)\nonumber\\
    & & {} + \frac{3g_2^4 + 6g_2^2 g_1^2 +
    3g_1^4}{1024\pi^2}\phi^4\left(log\frac{(g_2^2 +
    g_1^2)\phi^2}{4M^2}- \frac{5}{6} \right)\nonumber\\
    & & {} -\frac{3g_t^4\phi^4}{64\pi^2} \left(log
    \frac{g_t^2 \phi^2}{2M^2} - \frac{3}{2}
    \right)\nonumber\\
    & & {} + \frac{9
    \lambda^2\phi^4}{256\pi^2}\left(log\frac{3 \lambda
    \phi^2}{2 M^2} - \frac{3}{2} \right) \nonumber\\
    & & {} + \frac{3 \lambda^2 \phi^4}{256\pi^2}\left(
    log\frac{\lambda\phi^2}{2M^2} - \frac{3}{2} \right) +  \Omega
\end{eqnarray}

We will work with the form (23). We plug equation (23) into
equations (24) - (26) and obtain relations that embody REWSB
signatures:
\begin{equation}\label{eq: ndili17a}
 \frac{dV(\phi)}{d\phi}|_{\phi = M} =  \lambda - \frac{44}{3} C  = 0
\end{equation}
Putting  C value from equation (22) we obtain a quadratic equation
for $\lambda$:
\begin{equation}\label{eq: ndili17b}
\lambda^2 -\left(\frac{4 \pi^2}{11}\right) \lambda -
\frac{g_t^4}{4} + \frac{3g_2^4 + 2g_2^2 g_1^2 + g_1^4}{64} = 0
\end{equation}
Using the core coupling values from equation (7), the quadratic
equation becomes:
\begin{equation}\label{eq: ndili17c}
\lambda^2 - 3.588941 \lambda - \frac{g_t^4}{4} + 0.0105973 = 0
\end{equation}
This gives two possible values of $\lambda$ in terms of top quark
Yukawa coupling constant $g_t$, both at the renormalization scale
M = $\langle \phi \rangle = v/\sqrt 2$:
\begin{equation}\label{eq: ndili17d}
\lambda = 1.794470496 + \sqrt{3.204151685 + \frac{g_t^4}{4}}
\end{equation}
and
\begin{equation}\label{eq: ndili17e}
\lambda = 1.794470496 - \sqrt{3.204151685 + \frac{g_t^4}{4}}
\end{equation}
If we require that both coupling constants are positive : $\lambda
> 0; g_t > 0,$ then solution (31) implies that $\lambda \gg g_t $
with $\lambda = 3.584485$ at $g_t = 0$.  On the other hand, based
on equation (32), we find at $g_t = 0, \lambda = 0.004456,$ while
for $\lambda $ decreases to zero, $g_t = 0.50276 $. This solution
would indicate  that the scalar dynamics in the EWSB has a
coupling constant $\lambda$ that lies between zero and a value
0.004456 while the  Yukawa top quark coupling has a value that
lies between zero and 0.502758496. We may take the mean of each
parameter as its  value at the EWSB scale in this case of solution
(32). Then analogous to equation (7) we can write the $\lambda $
and $g_t$ values  at the EWSB scale for solution (32) as :
\begin{equation}\label{eq: ndili17f}
\lambda =  0.002278 ;  \frac{\lambda}{4 \pi^2} =  5.77 \times
10^{-5}; g_t = 0.25138 ; \frac{g_t^2}{4 \pi^2} = 5.07 \times
10^{-4}.
\end{equation}
In this way, we may claim that solution (32) represents a
situation where $g_t > \lambda $ at EWSB, while solution (31)
represents a situation where $ \lambda \gg g_t $ at EWSB scale.
These two situations we can also try to  distinguish in our
signature analysis of REWSB. \\

In the case of solution (31), we can use the known top quark mass
of 175 GeV  along side  equation (8) to estimate $g_t$ at EWSB
scale.  We get that $g_t = 1.00605,$ giving a corresponding value
of $\lambda$ from equation (31):
\begin{equation}\label{eq: ndili17g}
\lambda =  3.65464602 ; \frac{\lambda}{4 \pi^2} = 0.092573 ; g_t =
1.00605; \frac{g_t^2}{4 \pi^2} = 0.025637
\end{equation}

As seen from equations (33) and (34) both solutions (31) and (32)
are perturbative, if we define  perturbativity condition as:
\begin{equation}\label{eq: ndili17h}
\frac{\lambda}{4 \pi^2}  <  1  ;   \frac{g_t^2}{4 \pi^2}  < 1.
\end{equation}
The two solutions differ however in a profound way. This is that
if the dynamics of EWSB is very weakly perturbative (both
$\lambda$ and $h_t$ very small) as per equations (32) and (33),
REWSB predicts that the top quark coupling is relatively much
stronger than the scalar self interaction $\lambda$ at the EWSB
scale. On the other hand, if the EWSB dynamics is more strongly
coupled, (both $\lambda$ and $g_t$ large),  though still
perturbative as per equations (31) and (34), the scalar self
coupling constant $\lambda$ is relatively much larger than the top
quark Yukawa coupling constant $g_t$. These signatures of the
REWSB can be ascertained experimentally. \\

We can go further to analyze equations (25) and (23) for Higgs
mass. We obtain:
\begin{equation}\label{eq: ndili1b}
\frac{d^2V(\phi)}{d\phi^2} |_{\phi = M}  = 3 \lambda v^2 - 36Cv^2
= m_{h}^2
\end{equation}
Using  $\lambda = (44/3)C $ from equation (28), this becomes:
\begin{equation}\label{eq: ndili1c}
 m_{h}^2 = 8 v^2 C = 8 v^2 \left[
               \frac{12 \lambda^2 - 3 g_t^4}{64\pi^2} +
               \frac{3(3g_2^4 + 2g_2^2 g_1^2 + g_1^4)}{1024 \pi^2}
               \right]
\end{equation}

Using core values from equation (7) we can rewrite equation (37)
as:
\begin{equation}\label{eq: ndili1ca}
 m_{h}^2 = \frac{3 v^2}{8 \pi^2} \left[
               4 \lambda^2 -  g_t^4 +
 0.042389076 \right]
\end{equation}
Next we use equation (31) or (32) to eliminate $\lambda$. We
obtain:
\begin{equation}\label{eq: ndili1d}
 m_{h}^2 = \frac{3 v^2}{8 \pi^2}\left[4\left( 1.794470496 \pm
 \sqrt{3.204151685 + \frac{g_t^4}{4}} \right)^2 - g_t^4 +
 0.042389076 \right]
\end{equation}
Finally using core equations (7) and (8)  we obtain a Higgs boson
mass for the REWSB model :
\begin{equation}\label{eq: ndili1d}
 m_{h}^2 = \frac{3 v^2}{8 \pi^2}\left[4\left( 1.794470496 \pm
 \sqrt{3.204151685 + \frac{m_t^4}{v^4}} \right)^2 - \frac{4m_t^4}{v^4} +
 0.042389076 \right]
\end{equation}
Taking a top quark mass  $m_t = 175 GeV$  this evaluates to $m_h =
347.388GeV,$ (the positive value). This appears high.\\

\section{ Renormalization group improved  Effective potentials.}
The Coleman-Weinberg 1-loop effective potential equation (23) used
in the above analysis of REWSB, has the feature that it was
defined at one and only one arbitrarily chosen mass scale point M
which is the renormalization scale of the otherwise divergent
effective potential.  This one scale point M had also to be chosen
as the EWSB point  $\langle \phi \rangle = v/\sqrt 2 = M $, with
$v = 246 GeV.$ This restricted choice of $\phi(x)$ was
necessitated by the need for the log terms in equations (23) - (27
) to vanish or remain small and perturbative. This feature of the
Coleman-Weinberg potential equation (23) remains even if we make
another arbitrary choice of the renormalization point M'. We
remain bound  to study our effective potential and its signature
values, only for field values $\phi$  in the vicinity of an
arbitrarily chosen renormalization  scale M.  In effect,
$V_{eff}(\phi(x))$ cannot be studied generally as a function of
Higgs field $\phi(x)$ to ascertain how the potential behaves in
these other regions of larger or smaller $\phi$, and to what
extent any new features such as new minima or maxima of
$V_{eff}(\phi)$ found in those regions, affect or modify the EWSB
features or signatures of $V_{eff}$ found from earlier limited
observation at one arbitrary renormalization point M. A way out of
the problem is to free $V_{eff}$ of its dependence on the
renormalization scale M. That is we make $V_{eff}$ a physical
observable, whose field $\phi$ and coupling constant parameters
$\lambda_i$ can now vary widely over any desired domains,
unaffected by any one arbitrarily chosen renormalization point M.
Of course this freedom of the parameters of $V_{eff}$ to vary
widely still calls for some restraint such that the perturbative
character of the overall loop expansion equations (15) and (16)
can still be guaranteed. Both objectives and the automatic summing
up of leading logarithms, are achieved by making the entire
effective potential equation (15) or (16), satisfy a
Renormalization Group equation(RGE)  given by:
\begin{equation}\label{eq: ndili24}
\left[M\frac{\partial}{\partial M}  + \beta_{\lambda}
\frac{\partial}{\partial \lambda} + \sum_i \beta_i
\frac{\partial}{\partial g_i} -\gamma \phi
\frac{\partial}{\partial \phi}\right] V_{eff}(M, \lambda, g_i ,
\phi..) =0
\end{equation}
Here $\beta_\lambda , \beta_i $, and $\gamma,$ are the RG
functions (the beta functions and the anomalous  dimension $\gamma
$ of the Higgs field).  This RGE has a standard solution we write
in the form :
\begin{equation}\label{eq: ndili25}
V_{eff}^{RG} = V(M(t), \lambda(t), g_i(t), \phi(t)...)
\end{equation}
and call the RG improved effective potential. All its parameters
$M(t), \lambda(t), g_i(t), \phi(t) ...,$  vary or run  widely,
each in a coordinated manner prescribed by its beta or gamma
function.  The variable $t$ parameterizes the arbitrary
renormalization scale, while the parameters of $V_{eff}^{RG}$  run
according to the equations:
\begin{equation}\label{eq: ndili26}
M(t) = M_o e^t
\end{equation}
\begin{equation}\label{eq: ndili27}
\phi(t) = \phi_o \xi(t) = \phi_o exp\left(- \int_0^t
\gamma(\lambda(t');  g_i(t'))dt'\right)
\end{equation}
with:
\begin{equation}\label{eq: ndili29a}
\gamma(t') = \sum_{n=1}^\infty \gamma^{(n)}(t').
\end{equation}
Also:
\begin{equation}\label{eq: ndili28}
\frac{d \lambda(t)}{dt} =  \beta_\lambda (\lambda(t) ; g_i(t)) =
\sum_{n=1}^\infty \beta_\lambda^{(n)}.
\end{equation}
\begin{equation}\label{eq: ndili29}
\frac{d g_i(t)}{dt} =  \beta_{g_i}(g_i(t) ; \lambda) =
\sum_{n=1}^\infty \beta_{g_i}^{(n)}.
\end{equation}

 In the above equations,  $M_o$ is some fixed initial
renormalization point or reference scale,  which we shall choose
as $M_o = M_Z = 91.2GeV$, the $Z^o$ gauge boson mass. Similarly
the value of the Higgs field  $\phi_o$ at $t = 0$ we shall choose
as the field with VEV $ \langle \phi_o \rangle = v/ \sqrt 2$ where
$v = 246 $ GeV. The beta and gamma functions have to be calculated
separately to some loop order, and plugged into the above running
equations. Explicit expressions of these $\beta_i$ and $\gamma$
functions computed to 1-loop and 2-loop orders for the EWSB
system, have been tabulated by Ford et al. [8]. We shall rely on
their values to compute numerically the parameters
$\xi(t), \lambda(t), g_i(t)$ defined in equations (43)-(47). \\

In order to solve these integral and differential equations, we
require to specify the boundary conditions of the  equations.
These boundary conditions are the values of the parameters at some
chosen scale we specify as $t = 0$ which means scale $ M = M_0 =
M_Z = 91.2$GeV. Our explicit values of these parameters at $t = 0$
we take from equations (7), (33) and (34). \\

Once we have computed the explicit details of how the parameters
vary with $t$ over a wide range of values in equations (43) -
(47), our aim will be to use the new RG improved effective
potential equation (42) to re-examine the issue of signatures of
REWSB. To proceed with the analysis, we need first to expand
equation (42) into perturbative loop series and truncate the
series at a convenient loop order chosen here as 1-loop order.
Thus by analogy with equations (16) and (20), we  write our RG
improved effective potential equation (42) as:
\begin{equation}\label{eq: ndili30}
V_{eff}^{RG}(\phi) = V_o^{RG}(\phi)  +  V_{1L}^{RG}(\phi)  +
V_{2L}^{RG} (\phi) + .....
\end{equation}
To 1-loop order we write the RG improved REWSB  potential as:
\begin{equation}\label{eq: ndili32}
V_{eff}^{RG} = \frac{\lambda(t) \phi^4(t)}{4} + \phi^4(t)
\left[\frac{12 \lambda^2(t) - 3 g_t^4(t)}{64\pi^2} +
               \frac{3(3g_2^4(t) + 2g_2^2(t) g_1^2(t) + g_1^4(t))}{1024 \pi^2}
               \right] (log\frac{\phi^2(t)}{M^2(t)} - \frac{25}{6})
\end{equation}
which replaces equation (23). Similarly, corresponding to equation
(27) we would now write the RG improved 1-loop potential of Ford
et. al. [8]  and Casas et. al. [9-10]:
\begin{eqnarray}\label{eq: ndili33}
V_{eff}^{RG} &=& \frac{6g_2^4(t)\phi^4(t)}{1024\pi^2} \left(
log\frac{g_2^2(t)\phi^2(t)}{4M^2} - \frac{5}{6} \right)\nonumber\\
    & & {} + \frac{3g_2^4(t) + 6g_2^2(t)g_1^2(t) +
    3g_1^4(t)}{1024\pi^2}\phi^4(t)\left(log\frac{(g_2^2(t) +
    g_1^2(t))\phi^2(t)}{4M^2(t)}- \frac{5}{6} \right)\nonumber\\
    & & {} -\frac{3g_t^4(t)\phi^4(t)}{64\pi^2} \left(log
    \frac{g_t^2(t) \phi^2(t)}{2M^2(t)} - \frac{3}{2}
    \right)\nonumber\\
    & & {} + \frac{9
    \lambda^2(t)\phi^4(t)}{256\pi^2}\left(log\frac{3 \lambda(t)
    \phi^2(t)}{2 M^2(t)} - \frac{3}{2} \right) \nonumber\\
    & & {} + \frac{3 \lambda^2(t) \phi^4(t)}{256\pi^2}\left(
    log\frac{\lambda(t)\phi^2(t)}{2M^2(t)} - \frac{3}{2} \right).
\end{eqnarray}
We will work with equation (49), and comment  on equation (50)
later. Our analysis of equation (49) now proceeds in stages as
follows. \\

First we note that by construction the full (untruncated) RG
improved effective potential equation (42) or (48)  becomes
independent of scale, satisfying :
\begin{equation}\label{eq: ndili25a}
\frac{d V_{eff}^{RG}(M(t), \lambda(t), g_i(t), \phi(t)...)}{d t} =
0
\end{equation}
This  scale invariance applies also  to all order  derivatives of
$V_{eff}^{RG}$ with respect to any of its parameters such as
$\phi$. Thus we also have the conditions:
\begin{equation}\label{eq: ndili25b}
\frac{d}{dt}  \left(\frac{\partial V_{eff}^{RG} (M(t), \lambda(t),
g_i(t), \phi(t)...)}{\partial \phi(t)}\right) = 0
\end{equation}
and:
\begin{equation}\label{eq: ndili25b}
\frac{d}{dt}  \left(\frac{\partial^2 V_{eff}^{RG} (M(t),
\lambda(t), g_i(t), \phi(t)...)}{\partial^2 \phi(t)}\right) = 0
\end{equation}
The implication of these equations is that equations (24)-(26)
from which we determine the state of occurrence of SSB and the
resulting Higgs mass as well as  signatures of REWSB  (all arising
 from inherent dynamics of the system), have become independent of
 any one choice of renormalization scale t or M(t) at which we
 evaluate the system.
In a way, $V_{eff}^{RG}$ is now defined simultaneously over a
whole range of the renormalization scale from $t = 0 $ to
$\infty,$ at any one point of which we can freely compute the
dynamical quantities in equations (24) - (26) and determine at
what points we have SSB and its corresponding mass spectra,
unaffected by the value of the renormalization scale t at that
point, except that the results obtained at any one  t scale or
point, differ from the results at another t scale or point  by
mere change or transformation of  scale, but not intrinsic
difference of dynamics of the system. \\

 We note however that the above discussion regarding scale
 invariance of the RG improved effective potential holds
strictly  only where the potential is not truncated as in
equations (49) and (50). Where  $V_{eff}^{RG}$ is truncated and
its arguments are evaluated with beta and gamma functions that are
also computed only to a finite order (1-loop , 2-loop,... beta and
gamma functions), equations (51)-(53) cease to hold exact. We
correct for this  by looking for a restricted stretch of the t
scale, call it $t^*$ with $t_1 \le t^* \le t_2 $,  where the
truncated functions in equations (49) - (53) may still be
considered reasonably flat or minimally dependent on t. We  work
with such flat regions for potentials (49)- (53), in place of the
region  $t = (0, \infty)$ of the exact scale invariant
quantities. Now we analyze equations (49), (24)-(26). \\

 Our first step is to compute the various running parameters of
 $ V_{eff}^{RG},$ specified in equations (43) - (47).
  We illustrate our computational procedures by
 first considering the simpler case of 1-loop beta and
gamma functions. Later we treat the case of 2-loop beta and gamma
functions. To 1-loop beta and gamma functions [8], the running parameter
 equations to  compute are: \\
\begin{equation}\label{eq: ndili30a}
g_1^2(t) = \frac{g_{01}^2}{1 - (41/48 \pi^2) g_{01}^2 t }
\end{equation}
\begin{equation}\label{eq: ndili30b}
g_2^2(t) = \frac{g_{02}^2}{1 + (19/48 \pi^2) g_{02}^2 t }
\end{equation}
\begin{equation}\label{eq: ndili30c}
g_3^2(t) = \frac{g_{03}^2}{1 + (7/8 \pi^2) g_{03}^2 t }
\end{equation}
with $g_{01} = 0.3407; g_{02} = 0.6585 $ from equation (7) and $ g_{03} = 1.23. $\\
Also we get values of $g_t(t), \lambda(t), \xi(t)$ by numerically
integrating the following equations:
\begin{equation}\label{eq: ndili30d}
t =  16 \pi^2 \int_{g_{t0}}^{g_t(t)} \frac{dg_t}{\frac{9}{2}g_t^3
- g_t[8 g_3^2 + \frac{9}{4} g_2^2  + \frac{17}{12} g_1^2 ]}
\end{equation}
\begin{equation}\label{eq: ndili30e}
t =  16 \pi^2 \int_{\lambda_0}^{\lambda(t)} \frac{d \lambda(t)}{4
\lambda^2 + B \lambda + C}
\end{equation}
where $ B  = 12g_t^2 - 9 g_2^2 - 3 g_1^2$ ; and  $C = -36g_t^4 +
(9/4) g_1^4 + (9/2) g_2^2 g_1^2 + (27/4) g_2^4. $ Here we  use as
boundary values, the pair of values ($\lambda_0 = 3.65464602 ;
g_{t0} = 1.00605$) from equation (34). Later we shall consider the
other pair of values ($\lambda_0 = 0.002278 ; g_{t0} = 0.2513$) of
equation (33). Finally we have for the $\phi(t) $ and $\xi(t)$ :
\begin{equation}\label{eq: ndili30f}
\phi(t) = \phi_0 \xi(t) = \phi_0 exp \left(- \frac{1}{16
\pi^2}\int_0^t [3 g_t^2 - \frac{9}{4} g_2^2 - \frac{3}{4}
g_1^2]dt'\right)
\end{equation}
with this integral separated into two parts for purposes of
numerical computation :
\begin{equation}\label{eq: ndili30g}
\frac{1}{16 \pi^2}\int_0^t [3 g_t^2 - \frac{9}{4} g_2^2 -
\frac{3}{4} g_1^2]dt' = \frac{1}{16 \pi^2}\int_{g_{t0}}^{g_t(t)}[3
g_t^2]dg_t(t') - \frac{1}{16 \pi^2}\int_0^t [\frac{9}{4} g_2^2
+\frac{3}{4} g_1^2]dt'
\end{equation}
We proceeded by assigning only positive values to scale t, and
insisting in our numerical computations of equations (54) - (60)
that we obtain  only matching positive values of the coupling
constants  $g_1(t), g_2(t), g_3(t), g_t(t)$, and $\lambda(t)$. The
results we obtained are shown in \textbf{figure 1}. The potential
$V_{eff}^{RG}$ equation (49) is next plotted against t, to search
for a scale region $t^*$ where $V_{eff}^{RG}$ is reasonably flat,
independent of t. From the plot  shown in \textbf{fig. 2} we see
that equation (49) is reasonably scale invariant in the region: $0
\le t^* \le 8 $ . This allows us to compute the quantities:
$\partial V^{RG} /\partial \phi(t)$ and $\partial^2 V^{RG} /
\partial^2 \phi(t)$  at any point within this flat $t^*$ region,
and to run  the quantities down to our reference scale $t = 0$. It
is in this way that we obtain below, a  Higgs (running) mass at
the electroweak scale $t = 0$.  The relevant quantities to compute
and run in region $t^8$ are:
\begin{equation}\label{eq: ndili42a}
\frac{\partial V_{eff}^{RG}}{\partial \phi(t)} =  0
\end{equation}
implying  a SSB condition any where in region $t^*$ given by :
\begin{equation}\label{eq: ndili30e}
\lambda(t^*) + 4H(t^*) \left(log\frac{\phi^2(t^*)}{M^2(t^*)} -
\frac{25}{6} \right) + 2 H(t^*) = 0
\end{equation}
where :
\begin{equation}\label{eq: ndili30h}
H(t^*) = \left[\frac{12 \lambda^2(t^*) - 3 g_t^4(t^*)}{64\pi^2} +
               \frac{3(3g_2^4(t^*) + 2g_2^2(t^*) g_1^2(t^*) + g_1^4(t^*))}{1024 \pi^2}
               \right]
\end{equation}
Also :
\begin{equation}\label{eq: ndili42b}
\frac{\partial^2 V_{eff}^{RG}}{\partial^2 \phi(t)} =  m_h^2(t)
\end{equation}
which combined with equation (62) gives within region $t^*$, a
running Higgs mass of :
\begin{equation}\label{eq: ndili42c}
m_h^2(t^*) = 8 \phi^2(t^*) H(t^*)
\end{equation}
A plot of this $m_h^2(t^*) $ is shown in \textbf{figure 2} and
yields a Higgs (running) mass value of $m_h = 245.6 $ GeV at the
electroweak scale t = $t^*$ = 0. The corresponding physical Higgs
mass differs from this by only a small vacuum polarization
correction discussed by Casas et. al. [9,10] We will not go into this. \\

Based on \textbf{figure 2}, we can interpret the region $t \geq
9.0$ not only as a region beyond which the 1-loop  truncated RG
improved effective potential is not reliable, but also as a cut
off point $\Lambda$ beyond which the Standard model EWSB theory is
probably not valid. This will lead us to say that radiative
electroweak symmetry breaking (REWSB) with RG improved  effective
potential (at 1-loop beta function) predicts a Higgs mass around
245.6 GeV, as well as a cut off energy scale $\Lambda  \approx 8.0
\times 10^5 $ GeV  or less. \\ \\

\section{The Case $g_t > \lambda$ at electroweak scale}
We examine next the features of REWSB in the case  $g_t
> \lambda $ at electroweak scale, presented  by
 equation (33), in place of equation (34). We recall that
equations (33) and (34) provided us a means to discriminate
between  two scenarios where at EWSB scale, the scalar field
coupling $\lambda$ is stronger (equation (34)) or weaker (equation
(33)) than the top quark Yukawa coupling. Our calculations in
sections 4 and 5, dealt with the case $\lambda_0 > g_{t0}$. Here
we re-compute  the same RG improved effective potential equation
(49) under this alternative solution equation (33). Specifically
we take as our new boundary values, the values given in equation
(33): $\lambda_o = 0.002278 $ and  $ g_{t0} = 0.25138 $.
 We then re-compute our running parameters and the effective potential (49),
 together with the running Higgs mass. The results and features we
 find with equation (33) and the 1-loop beta and gamma function equations
(43) - (47) are shown in \textbf{figure 3 }. They lead us to a
running Higgs mass of only 6.65GeV shown in \textbf{figure 4}.
Since experiments already rule out such low  mass Higgs boson, we
can draw  one conclusion that the REWSB  model clearly selects
only the solution (34): $\lambda_0 \gg g_{t0}$ at EWSB scale.  \\

\section{Case of 2-loop beta and gamma functions}
Working now with only solution (34), we consider finally the case
of 2-loop beta and gamma function RG improvement of the effective
potential equation (49). According to Kastening [12],Bando et. al
[13], one gets optimal improvement of a 1-loop effective potential
if we evaluate it using 2-loop beta and gamma functions. We test
this out  by re-computing our REWSB features and comparing with
our 1-loop results of sections 4 and 5. Accordingly, we replace
equations (54) to (60) by the following 2-loop beta function
running parameters:
\begin{equation}\label{eq: ndili60}
t =  16 \pi^2 \int_{g_{01}}^{g_1(t)} \frac{dg_1(t)}{g_1^3 \left[
\frac{41}{6} + \frac{199}{18} g_1^2 + \frac{9}{2} g_2^2 +
\frac{44}{3} g_3^2 - \frac{17}{6} g_t^2 \right ] }
\end{equation}
\begin{equation}\label{eq: ndili61}
t =  16 \pi^2 \int_{g_{02}}^{g_2(t)} \frac{dg_2(t)}{g_2^3 \left[
\frac{-19}{6} + \frac{3}{2} g_1^2 + \frac{35}{6} g_2^2 + 12 g_3^2
- \frac{3}{2} g_t^2 \right ] }
\end{equation}
\begin{equation}\label{eq: ndili62}
t =  16 \pi^2 \int_{g_{03}}^{g_3(t)} \frac{dg_3(t)}{g_3^3 \left[ -
7 + \frac{11}{6} g_1^2 + \frac{9}{2} g_2^2 - 26 g_3^2 - 2 g_t^2
\right ] }
\end{equation}
with $g_{01} = 0.3407; g_{02} = 0.6585 $ and $ g_{03} = 1.23$ as before.\\
For 2-loop $g_t(t)$ and $\lambda(t)$,we compute positive values of
the upper integration limits  $g(t)$ and $\lambda(t)$ that match
an assigned t value in the equations :
\begin{equation}\label{eq: ndili63}
t =  16 \pi^2 \int_{g_{t0}}^{g_t(t)} \frac{dg_t(t)}{-12 g_t^5(t) +
B_1 g_t^3(t) + C_1 g_t(t)}
\end{equation}
and :
\begin{equation}\label{eq: ndili64}
t =  16 \pi^2 \int_{\lambda_0}^{\lambda(t)} \frac{d
\lambda(t)}{-\frac{26}{3} \lambda^3(t) +  B_2 \lambda^2(t) + C_2
\lambda(t) + D}
\end{equation}
where :
\begin{eqnarray}
B_1 & = & \frac{9}{2} + \frac{131}{16}g_1^2 + \frac{225}{16}g_2^2
+ 36g_3^2 - 2\lambda \nonumber\\
C_1 & = & -8g_3^2 - \frac{9}{4}g_2^2 - \frac{17}{12}g_1^2 +
\frac{1187}{216}g_1^4 - \frac{3}{4}g_2^2 g_1^2
+\frac{19}{9}g_1^2g_3^2 - \frac{23}{4}g_2^4 + 9g_2^2g_3^2 -
108g_3^4 + \frac{1}{6}\lambda^2\nonumber\\
 B_2 & = & 4 - 24g_t^2 + 6(3g_2^2 + g_1^2) \nonumber\\
 C_2 & = & 12g_t^2 -9g_2^2 - 3g_1^2 - 3g_t^4 + 80g_3^2g_t^2 +
 \frac{45}{2}g_2^2g_t^2 + \frac{85}{6} g_1^2g_t^2 -
 \frac{73}{8}g_2^4 +\frac{39}{4}g_2^2g_1^2 +
 \frac{629}{24}g_1^4\nonumber\\
 D  & =  & -36g_t^4 + \frac{9}{4}g_1^4 + \frac{9}{2} g_2^2g_1^2 +
 \frac{27}{4}g_2^4 + 180g_t^6 - 192g_t^4g_3^2 - 16g_t^4g_1^2
 -\frac{27}{2}g_t^2g_2^4\nonumber\\
   &  & {} +63g_t^2g_2^2g_1^2 - \frac{57}{2}g_t^2 g_1^4 +
   \frac{915}{8}g_2^6 - \frac{289}{8}g_2^4g_1^2 - \frac{559}{8}
   g_2^2g_1^4 - \frac{379}{8}g_1^6
   \end{eqnarray}
 The boundary values we use are from equation (34):
 $\lambda_0 = 3.65464602$ and $ g_{t0} = 1.00605. $ Also because
 of the greatly intertwined nature of the 2-loop $\beta, \gamma$
 functions  and the running parameters, certain approximations had to
 be made in which  quantities like $B_1, B_2, C_1, C_2, D$, in the
 above integrands, were treated as constants having values
corresponding to a chosen t value, or else to their known 1-loop  values. \\

Finally  for $\phi(t) $ and $\xi(t)$ we have the 2-loop gamma
function equation:
\begin{equation}\label{eq: ndili65}
\phi(t) = \phi_0 \xi(t) = \phi_0 exp \left(- \frac{1}{16
\pi^2}\int_0^t [\gamma_1(t') + \gamma_2(t')]dt'\right)
\end{equation}
where:
\begin{eqnarray}
\gamma_1(t) & = & 3 g_t^2(t) - \frac{9}{4} g_2^2(t) - \frac{3}{4}
g_1^2(t)\nonumber \\
\gamma_2(t) & = & \frac{1}{6} \lambda^2 - \frac{27}{4}g_t^4 +
20g_3^2g_t^2 + \frac{45}{8}g_2^2g_t^2 + \frac{85}{24}g_1^2g_t^2 -
\frac{271}{32}g_2^4 + \frac{9}{16}g_2^2g_1^2 +
\frac{431}{96}g_1^4\nonumber\\
  &  &
\end{eqnarray}
The results of computing these 2-loop running  parameters, and
plugging into equations (49), (24) - (26), are shown in
\textbf{figures 5 and 6.} Some individual running parameters
appear greatly modified at the 2-loop level. This is particularly
so for $\lambda(t)$ and $g_2(t)$ seen from figures 1 and 6.
Correspondingly, the 2-loop RG improved effective potential   does
not exhibit much of a flat $t^*$ region seen in figure 6 compared
to 1-loop figure 2. The value of running Higgs mass at EWSB scale
:$t = 0$ predicted in both cases is however about the same : $m_h
= 245.6 $ GeV. \\

\section{ Relating our results to other calculations}
We can compare these features of our RG improved REWSB with
features found in conventional electroweak symmetry breaking
calculations. We already showed in section 3 that CEWSB  (at tree
level), does not predict any explicit value of the Higgs mass
$m_h$ nor the scalar coupling constant $\lambda$.  But the
conventional electroweak symmetry breaking potential (with $\mu^2
\ne 0)$ can also have higher order potential terms added to it as
in equation (16), and treated to a RG group improvement as for our
REWSB model considered in sections 5 and 6. This has been done by
a number of authors [7-17], particularly Ford et. al. [8] and
Casas et. al. [9,10]. We can look at these RG improved CEWSB
results and compare them to our REWSB results to see any
particular signature features. These authors [8-10]  worked with
1-loop effective potential  of the form shown in equations (27)
and (50) to which non-zero scalar field mass term $\mu^2 \ne 0$ is
added from equation (1). This RG improved CEWSB equation (50) is
then analyzed by the usual equations (24) - (26),  combined with
the additional  requirement of vacuum stability or $\lambda \geq
0$. Casas et. al [9,10] obtained  a Higgs mass estimate of $ m_h >
128 \pm 33 $GeV, to be compared with our explicit REWSB result of
running Higgs mass $m_h = 245.6 $ GeV.  Admittedly, many of these
CEWSB RG improved effective potential analysis, were done at a
time the top quark mass was still considered unknown free
parameter, in contrast  to our  treatment. Ford et. al. [8] for
example stated their result in terms of the top quark mass : $m_h
\geq 1.95 m_t - 189 $GeV. Overall, there appears a need to
re-analyze the RG improved CEWSB  effective potential case for
better  comparison with REWSB, and to ascertain
the role the scalar field mass parameter $\mu \ne 0$ actually  plays. \\

Another comparison we can make is with the recent work of Elias
et. al.[2-6] who analyzed the same radiative electroweak symmetry
breaking (REWSB), and  raised the same issue of the possible
experimentally accessible signature differences of REWSB and
CEWSB.  They worked not with the standard Coleman-Weinberg Feynman
loop calculated effective potential equations (15), (23), (27),
(49), (50), but took advantage of the fact that the loop
calculated effective potentials invariably involved leading
logarithms  as in equation (20), multiplied by a factor C that
involves products of EW coupling constants and the scalar field
$\phi$. Elias et. al. showed that this  usual Coleman-Weinberg RG
improved effective potential in the particular case of REWSB can
be realized as a direct summation to infinity of leading logarithm
terms. Putting such leading logarithm parameterized effective
potential through the usual equations (24) - (26), Elias et. al.
obtained a REWSB Higgs mass ranging in value from $m_h = 216 - 231
$ GeV.  They obtained  ssociated scalar field coupling $\lambda =
2.15 $. These results are comparable to our values $m_h = 245.6$
GeV, and $\lambda = 3.65$. Their values for other EW coupling
constants are also very comparable, Thus while for the top quark,
Elias et. al. obtained the value $g_{t0} = 1.00$ at EWSB scale, we
obtained in equation (34) the value $g_{t0} = 1.00605$.\\ \\

\section{Summary and Conclusions}.
We find close agreement between  our results based on the standard
Coleman-Weinberg loop calculated, and renormalization group
improved, effective potential, and the Elias et. al. results based
on direct summation of leading logarithm terms. Basing on both
calculations of the same radiative electroweak symmetry breaking
(REWSB), we can re-affirm  that REWSB model not only remains a
viable mechanism of electroweak symmetry breaking, but is able to
give an explicit value of the (running)  Higgs mass. The (running)
Higgs mass predicted by the REWSB mechanism, lies  in the range
215 - 250 GeV, at electroweak scale. In contrast the CEWSB models
of Ford et. al [8], Casas et al. [9.10] and others [7,14-16], are
only able to put a lower bound typically $m_h \geq 1.95 m_t - 189
$GeV on the Higgs mass. Because of this limited information from
CEWSB, we cannot use the Higgs mass value as a conclusive
signature to distinguish between  REWSB from CEWSB. \\

We are left to consider the coupling constants $\lambda$ and $g_t$
of the system.  Our calculations established two points regarding
$\lambda$ and $g_t$. We found that of the two possibilities  : $
\lambda > g_t$ or $g_t > \lambda$ at EWSB scale, the $\lambda \gg
g_t$ is the correct one, while the case $g_t > \lambda$ that
implies a Higgs mass $m_h \approx 6.65 GeV$ is already ruled out
by experiment. Explicitly, we obtained in our equation (34) that
the ratio $\lambda/g_t = 3.63$. Elias et. al. found  a lower
 ratio of 2.15. The two results are however consistent to the extent
 $\lambda \gg  g_t.$ \\ \\

 If we accept the REWSB value of Higgs mass of 215 - 250 GeV as
 physical Higgs mass holding whether the mechanism of EWSB is
 by REWSB or by CEWSB, we can deduce the corresponding $\lambda_{CEWSB}$ value
from equations (13) and (14). We find  $\lambda_{CEWSB} = 0.997$
to be compared with $\lambda_{REWSB} = 3.65 $ at the same Higgs
mass of 245.6 GeV.  We conclude as did Elias et. al. [2,3], that
the REWSB mechanism is associated with a greatly enhanced scalar
field coupling $\lambda_{REWSB} \gg \lambda_{CEWSB}$. \\

This  can become an important signature difference between REWSB
and CEWSB that can be verified experimentally.  The fact is that
based on equations (1) and (2), the individual  scalar field
components $\phi_1^+, \phi_2^+, \phi_4^o, \phi_3^o$  all couple
with the same strength, $\lambda_{CEWSB}$ or $\lambda_{REWSB}$
depending on the model.  When this fact is combined with the
equivalence theorem [18]  between EWSB Nambu-Goldstone bosons and
linearly polarized gauge bosons, one comes to the conclusion
 that interactions of $W_L^+ , W_L^- , Z_L^o$ can provide
 experimentally observable coupling constant based signature
 differentiation between REWSB and  CEWSB, a fact
 pointed out by Elias et. al. [2,3], but needing further study.   \\

The observed enhanced coupling  $\lambda_{REWSB} \gg
\lambda_{CEWSB}$ by itself calls for explanation. It suggests as
part of our main finding  that the scalar field couples less
strongly to achieve the same end result of EWSB in a quantum
system that has a pre-set mass scale $\mu^2 \ne 0 $, compared to
the same quantum system  with no pre-set scale $\mu^2 = 0$. The
same scalar field couples more strongly in the latter case.
Whether this is a general principle of quantum dynamics is
something  for further study.  \\ \\

\textbf{References :} \\

1. S. Coleman and E. Weinberg, Phys. Rev. D7 (1973) 1888 . \\

2. V. Elias, R.B. Mann, D.G.C.McKeon, and T.G. Steele, Phys. Rev.
Lett. 91, 251601. \\

3. F. A. Chishtie, V. Elias, and T. G. Steele, Intern. Journ. Mod.
Phys. A20 (2005) 6241. \\

4. V. Elias, R.B. Mann, D.G.C. Mckeon and T.G. Steele, Nucl. Phys.
B678 (2004) 147. \\

5. V. Elias,  R. B. Mann, D.G.C. McKeon and T.G. Steele,
arXiv.hep-ph/0508107 (2005). \\

6. F. A. Chistie,  V. Elias, R.B. Mann, D.G.C.McKeon, and T.G.
Steele, Nucl. Phys. B 743 (2006) 1034. \\

7. M. Sher, Phy. Report 179 (1989) 273. \\

8. C.Ford, D.R.T.Jones, P.W Stephenson and M B Einhorn, Nucl.
Phys.B395 (1993) 17. \\

9. J.A. Casas, J.R. Espinosa, M. Quiros and A. Riotto, Nucl.
Phys. B436 (1995) 3; (E) B439 (1995) 466. \\

10. J.A. Casas, J.R. Espinosa, M. Quiros, Phys. Lett. B342 (1995)
171. \\

11. T. P. Cheng and L. F. Li , Gauge theory of elementary particle
physics,  Oxford  University Press  (1984). \\

12. B. Kastening : Phys. Lett. B283 (1992) 287. \\

13. M. Bando, T. Kugo,  M. Maekawa and H. Nakano, Phys. Lett. B301
(1993) 83 ; Prog. Theor. Phys.  90 (1993) 405. \\

 14. G. Altarelli and G. Isidori Phys. Lett.  B 337 (1994) 141 \\

 15.M. Lindner, M. Sher and H.W. Zaglauer, Phys. Lett. B228 (1989)
 139. \\

 16. Vincenzo Branchina and  Hugo  Faivre, Phys. Rev.  D72  (2005)
 065017. \\

 17. P. Kielanowski,  S.R. Juarez W, and H.G. Solis-Rodriguez,
  Phys. Rev. D72 (2005) 096003). \\

18. P. N. Maher, L. Durand  and  K. Riesseimann, Phys. Rev.  D 48
(1993) 1061. \\

\textbf{Figure Captions: }

Figure 1 : Running parameters of RG improved effective potential
 at 1-loop $\beta$ and $\gamma$ functions  and for equation (34), $\lambda_o > g_{t0}$.   \\

Figure 2 : Flatness $t^*$ test and running Higgs mass  plot
yielding a running Higgs mass intercept of $m_h = 245.6 $ GeV at EW scale t = 0. \\

Figure 3 : Running parameters of RG improved effective potential
at 1-loop $\beta$  and $\gamma$ functions, with $g_{t0} > \lambda_o $,
equation (33) case.  \\

Figure 4  :  Flatness $t^*$ and running Higgs mass plots at 1-loop
$\beta$ and $\gamma$ functions, and for $g_{t0} > \lambda_o $ equation (33) case. \\

Figure 5   : Running parameters of RG improved effective potential
 at 2-loop $\beta$ and $\gamma$ functions  and for equation (34), $\lambda_o > g_{t0}$.   \\

Figure 6  :2-loop $\beta$ and $\gamma$ functions  flatness $t^*$
and running Higgs mass plots, yielding a running Higgs mass
intercept of  $m_h = 245.6 $ GeV at EW scale t = 0. \\

\begin{figure}
\scalebox{0.75}{\includegraphics{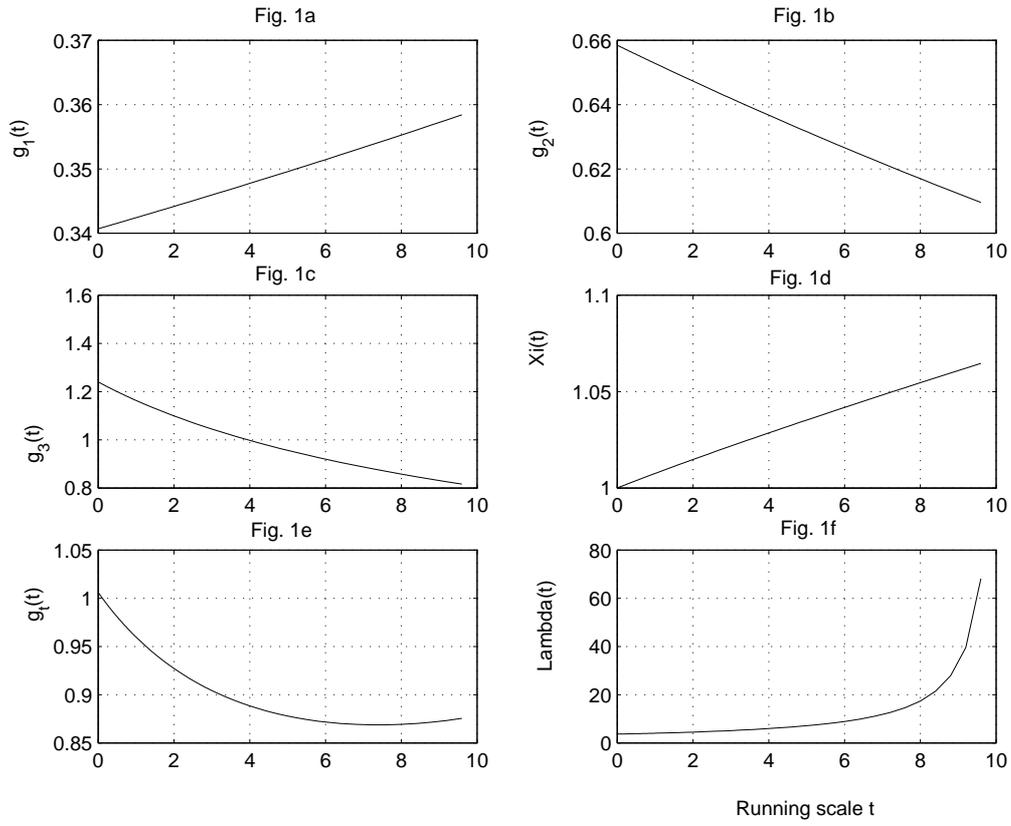}} \caption{Running
parameters of RG improved effective potential at 1-loop $\beta$
and $\gamma$ functions  and for equation (34), $\lambda_o
> g_{t0}$. } \label{fig: ndili29} \centering
\end{figure}

\begin{figure}
\scalebox{0.75}{\includegraphics{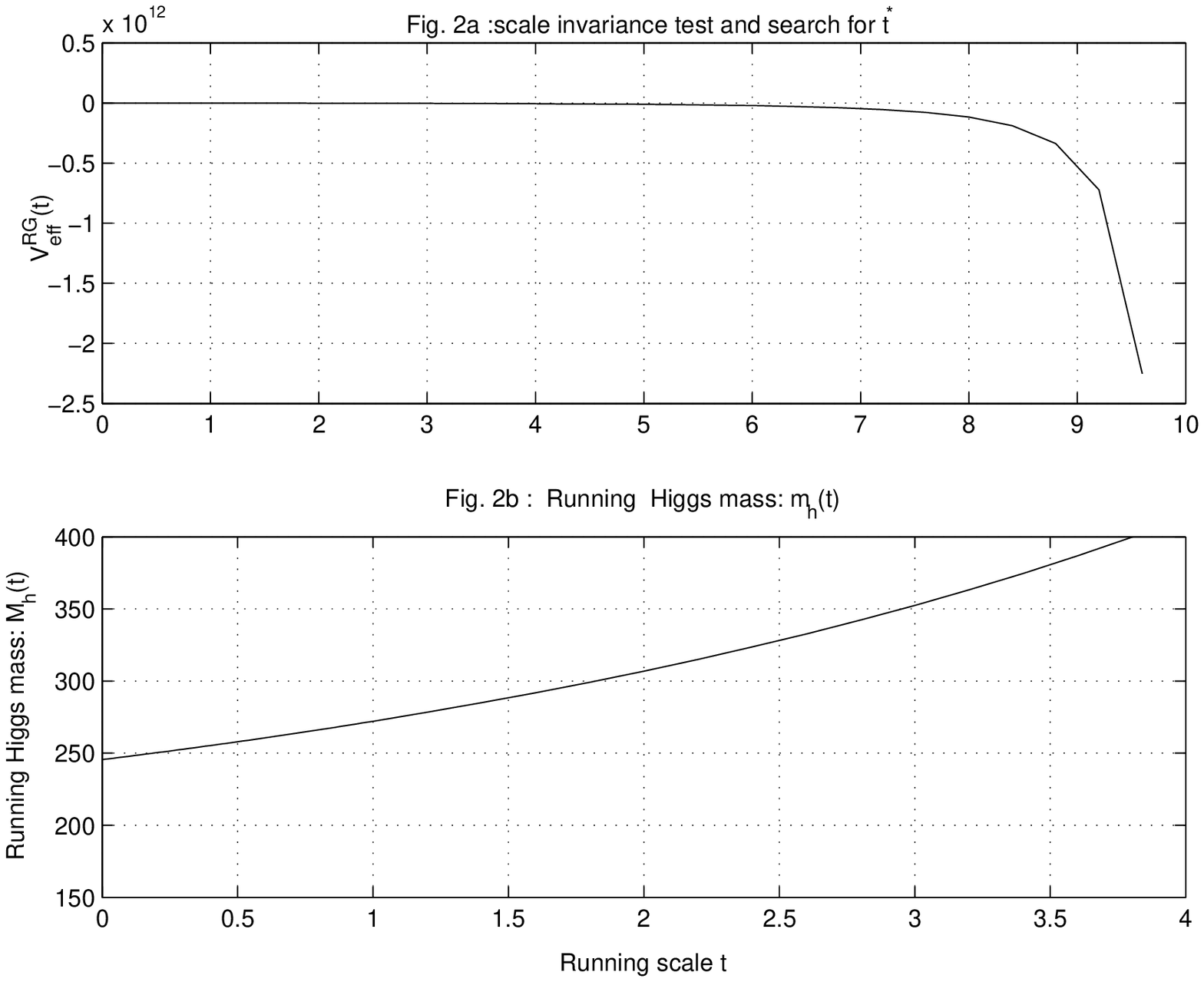}} \caption{Flatness
$t^*$ test and running Higgs mass  plot yielding a running Higgs
mass intercept of $m_h = 245.6 $ GeV at EW scale t = 0. }
\label{fig: ndili30} \centering
\end{figure}

\begin{figure}
\scalebox{0.75}{\includegraphics{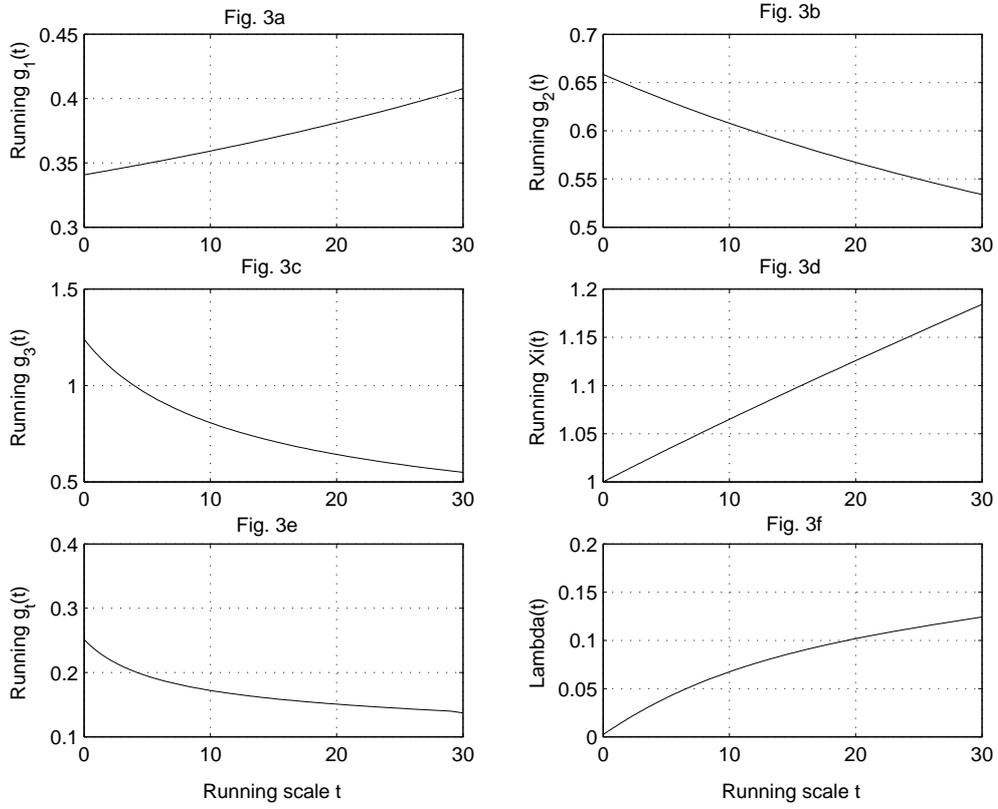}} \caption{Running
parameters of RG improved effective potential at 1-loop $\beta$
and $\gamma$ functions, with $g_{t0} > \lambda_o $, equation (33)
case.  } \label{fig: ndili31} \centering
\end{figure}

\begin{figure}
\scalebox{0.75}{\includegraphics{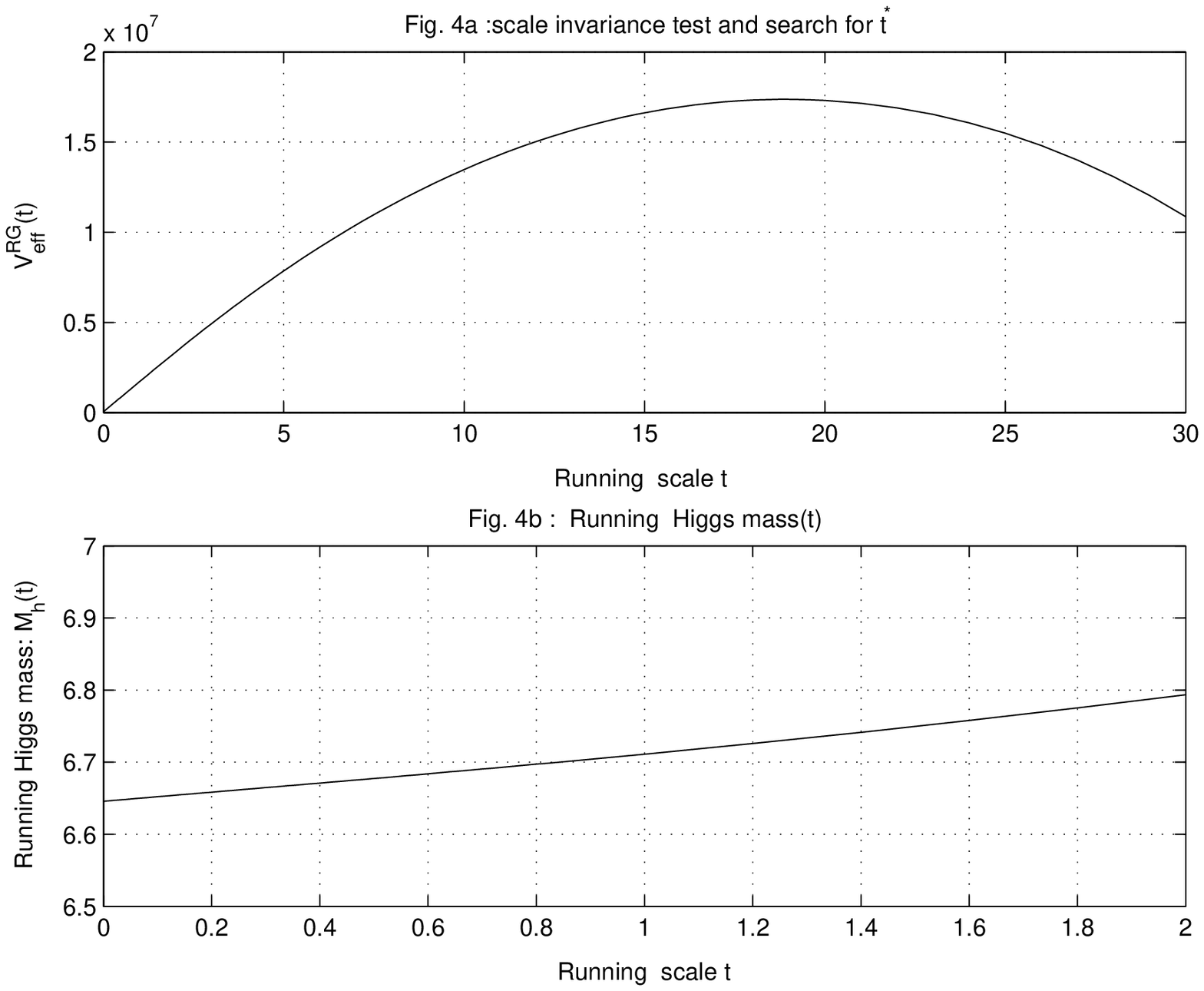}} \caption{ Flatness
$t^*$ and running Higgs mass plots at 1-loop $\beta$ and $\gamma$
functions, and for $g_{t0} > \lambda_o $ equation (33) case. }
\label{fig: ndili32} \centering
\end{figure}

\begin{figure}
\scalebox{0.75}{\includegraphics{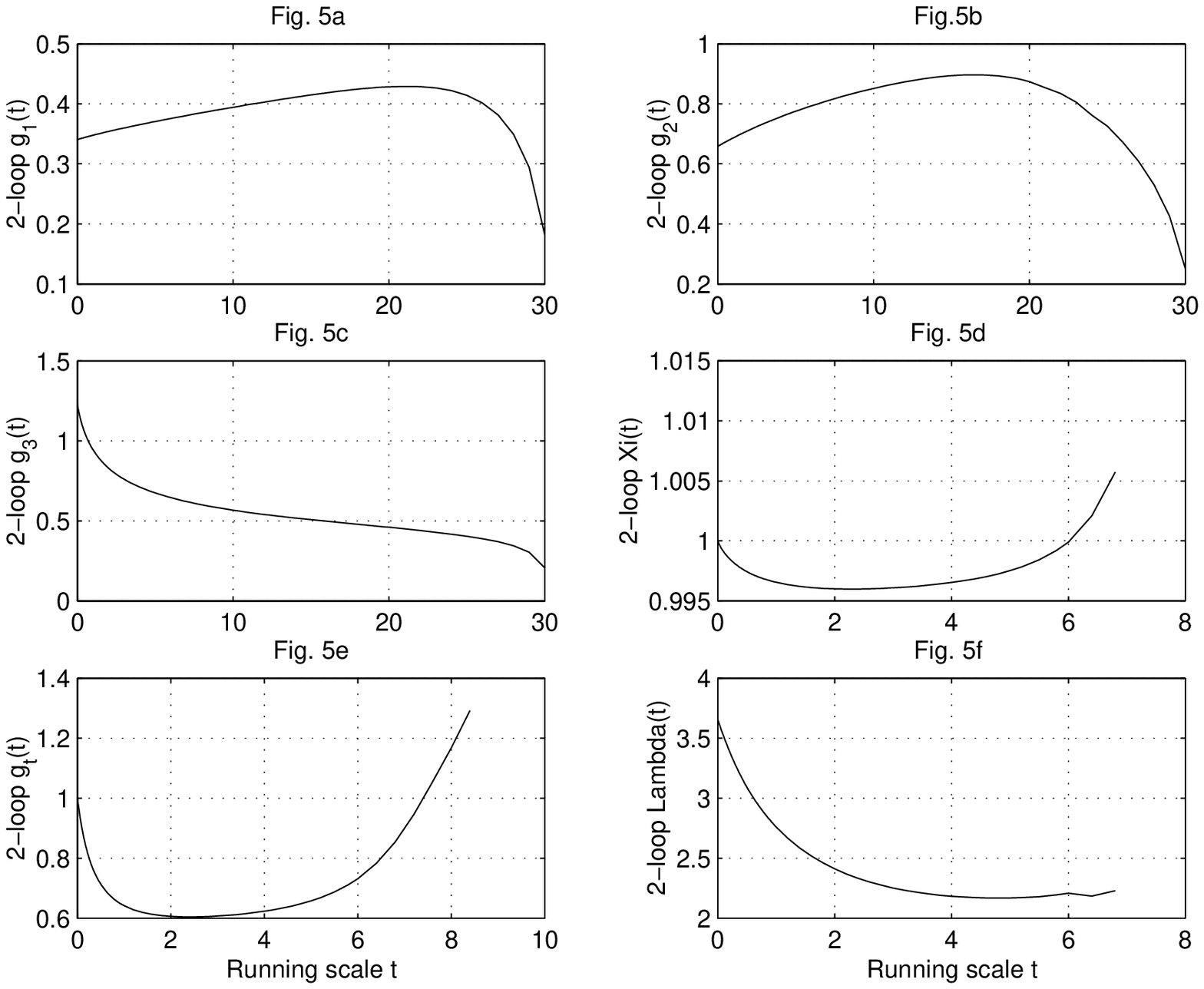}} \caption{Running
parameters of RG improved effective potential
 at 2-loop $\beta$ and $\gamma$ functions  and for equation (34), $\lambda_o > g_{t0}$. }
\label{fig: ndili33} \centering
\end{figure}

\begin{figure}
\scalebox{0.75}{\includegraphics{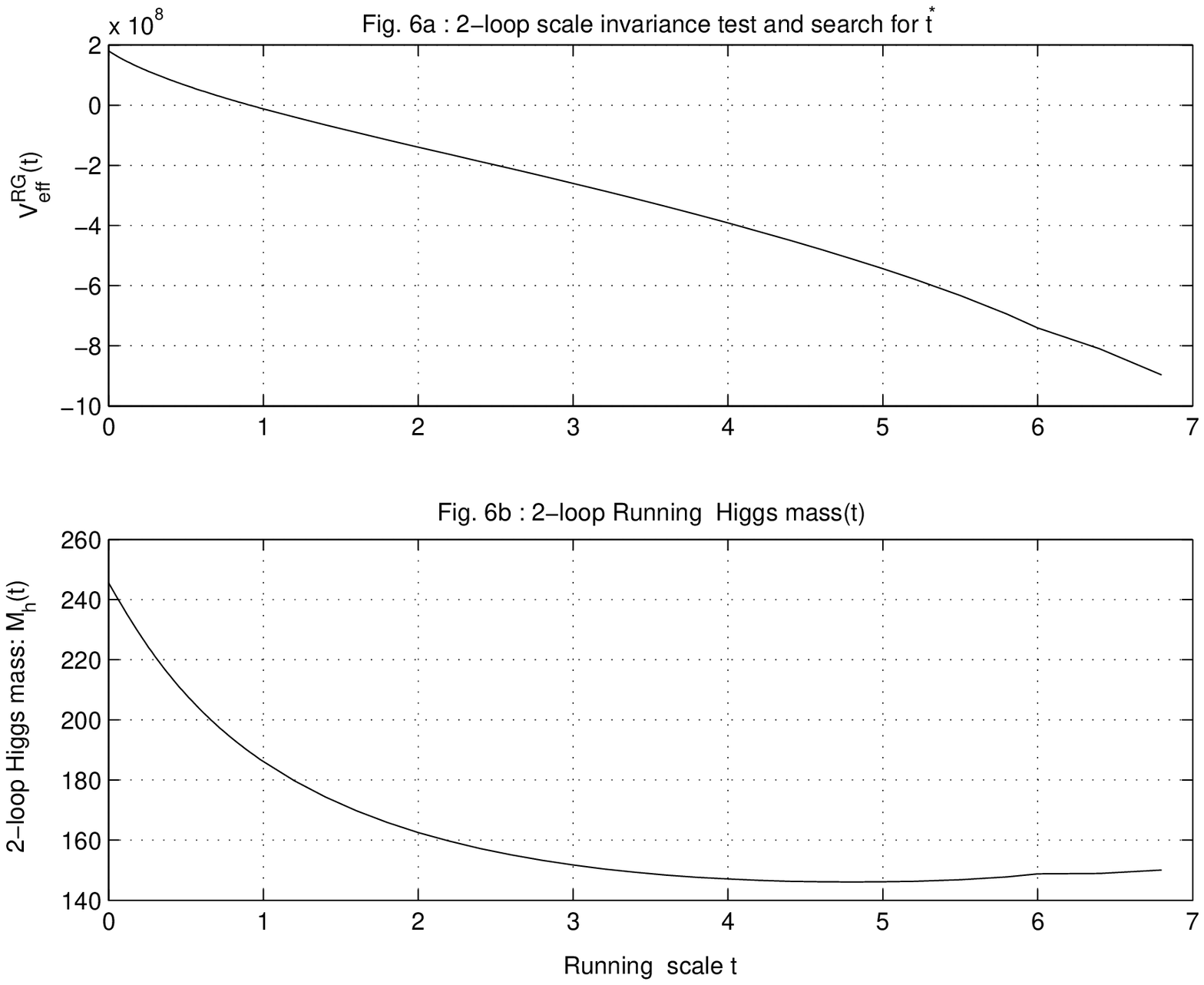}} \caption{2-loop
$\beta$ and $\gamma$ functions  flatness $t^*$ and running Higgs
mass plots, yielding a running Higgs mass intercept of  $m_h =
245.6 $ GeV at EW scale t = 0. } \label{fig: ndili34} \centering
\end{figure}

\end{document}